\documentclass[aps,prb,twocolumn,longbibliography,superscriptaddress]{revtex4-1}
\usepackage{epsfig}
\usepackage{epstopdf}
\usepackage{amsmath}
\usepackage{amsfonts}
\usepackage{amssymb}
\usepackage{hyperref}
\usepackage{bm}
\usepackage{makecell}
\usepackage{rotating}
\usepackage{hyperref}

\usepackage{graphicx}
\usepackage{dcolumn}
\usepackage{bm}
\usepackage{color}

\usepackage{tikz,xcolor,hyperref}

\definecolor{lime}{HTML}{A6CE39}
\DeclareRobustCommand{\orcidicon}{%
	\begin{tikzpicture}
	\draw[lime, fill=lime] (0,0)
	circle [radius=0.16]
	node[white] {{\fontfamily{qag}\selectfont \tiny ID}};
	\draw[white, fill=white] (-0.0625,0.095)
	circle [radius=0.007];
	\end{tikzpicture}
	\hspace{-2mm}
}

\foreach \x in {A, ..., Z}{%
	\expandafter\xdef\csname orcid\x\endcsname{\noexpand\href{https://orcid.org/\csname orcidauthor\x\endcsname}{\noexpand\orcidicon}}
}

\begin{document}

\title{Orbital-selective altermagnetism and correlation-enhanced spin-splitting \\ in transition metal oxides}

\author{Giuseppe Cuono\orcidD}
\email{gcuono@magtop.ifpan.edu.pl}
\affiliation{International Research Centre Magtop, Institute of Physics, Polish Academy of Sciences,
Aleja Lotnik\'ow 32/46, PL-02668 Warsaw, Poland}

\author{Raghottam M. Sattigeri\orcidC}
\affiliation{International Research Centre Magtop, Institute of Physics, Polish Academy of Sciences,
Aleja Lotnik\'ow 32/46, PL-02668 Warsaw, Poland}

\author{Jan Skolimowski\orcidE}
\affiliation{International Research Centre Magtop, Institute of Physics, Polish Academy of Sciences,
Aleja Lotnik\'ow 32/46, PL-02668 Warsaw, Poland}

\author{Carmine Autieri\orcidA}
\email{autieri@magtop.ifpan.edu.pl}
\affiliation{International Research Centre Magtop, Institute of Physics, Polish Academy of Sciences,
Aleja Lotnik\'ow 32/46, PL-02668 Warsaw, Poland}

\date{\today}
\begin{abstract}
We investigate the altermagnetic properties of strongly-correlated transition metal oxides considering the family of the quasi two-dimensional A$_2$BO$_4$ and three-dimensional ABO$_3$. 
As a test study, we analyze the Mott insulators Ca$_2$RuO$_4$ and YVO$_3$.
In both cases, the orbital physics is extremely relevant in the t$_{2g}$ subsector with the presence of an orbital-selective Mott physics in the first case and of a robust orbital-order in the second case.
Using \textit{first-principles} calculations, we show the presence of an orbital-selective altermagnetism in the case of Ca$_2$RuO$_4$.
In the case of YVO$_3$, we study the altermagnetism as a function of the magnetic ordering and of the Coulomb repulsion U. We find that the altermagnetism is present in all magnetic orders with the symmetries of the Brillouin zone depending on the magnetic order. Finally, the Coulomb repulsion enhances the non-relativistic spin-splitting making the strongly-correlated systems an exciting playground for the study of the altermagnetism. 
\end{abstract}


\maketitle

\section{Introduction}


The spin splitting of the energy bands typical for the ferromagnetic and ferrimagnetic configurations was found in antiferromagnetic systems, namely magnetically ordered compounds with zero net magnetization. These compounds with compensated magnetic moments in the real space but with spin-polarization order in the reciprocal space are named altermagnets\cite{Smejkal22,Smejkal22beyond}.
As a consequence of the non-relativistic spin splitting, the altermagnetic systems exhibit an anomalous Hall effect even without a magnetic order in the real space\cite{doi:10.1126/sciadv.aaz8809}. This happens when the opposite-spin sublattices are connected by roto-translations and not by inversion or translation mechanisms. These conditions are satisfied in many antiferromagnetic systems with antiferro orbital-order\cite{PhysRevLett.101.266405}.
The magnetic space groups are classified into four types, depending on the relation to the parent crystallographic space group\cite{bradley2010mathematical}.
In a recent work, it was shown that the altermagnetism can be found in the type-I and type-III magnetic space groups\cite{GUO2023100991}.
Several transition metal oxides\cite{GUO2023100991} and rare-earth compounds\cite{Cuono23EuCd2As2} have been shown to exhibit altermagnetism. We have shown that altermagnetism survives selectively on the surface states \cite{Sattigeri23} and has an interplay with the non-symmorphic symmetries\cite{Cuono19PRM,Campbell2021,Fakhredine23}. 
The altermagnets can also be used in applications like spintronics\cite{Shao21}, spincaloritronics\cite{zhou2023crystal} and in Josephson junctions\cite{Ouassou23}. 

\begin{figure}[t!]
\centering
\includegraphics[width=6.59cm,angle=0]{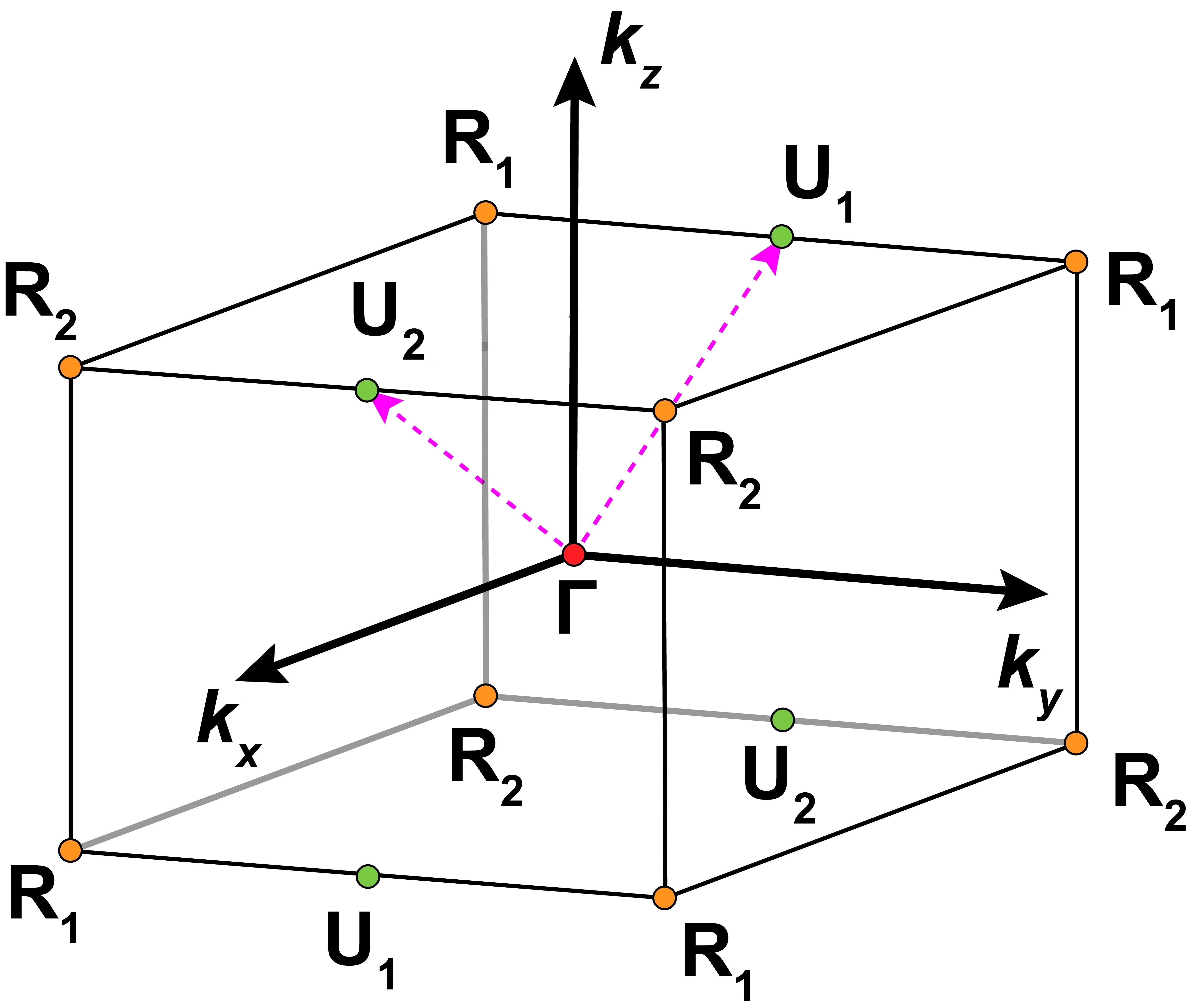}
\caption{Symmetries of the irreducible Brillouin zone for the orthorhombic Ca$_2$RuO$_4$ (space group Pbca no. 61). In our notation, the high-symmetry points with the subscript 1 and 2 show altermagnetism along the path towards the $\Gamma$ point. The position of the high-symmetry k-points U$_1$, U$_2$, R$_1$ and R$_2$ are highlighted in green and orange. The dashed magenta line indicates the high-symmetry path U$_1$-$\Gamma$-U$_2$ which is one of the possible paths to show the altermagnetism in this magnetic space group.}\label{Brillouin_CAO}
\end{figure}

\begin{figure}[t!]
\centering
\includegraphics[width=6.19cm,angle=270]{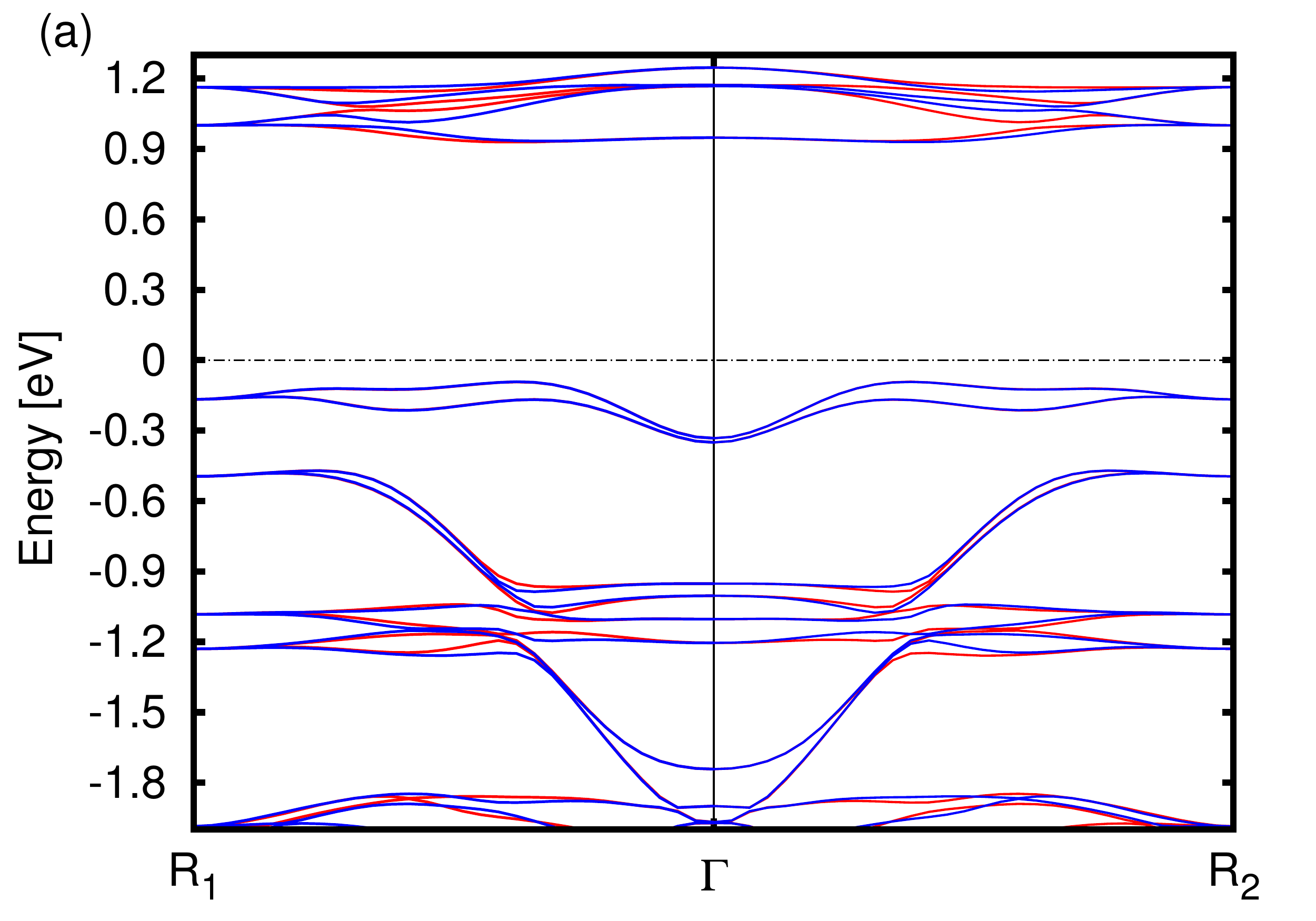}
\includegraphics[width=6.19cm,angle=270]{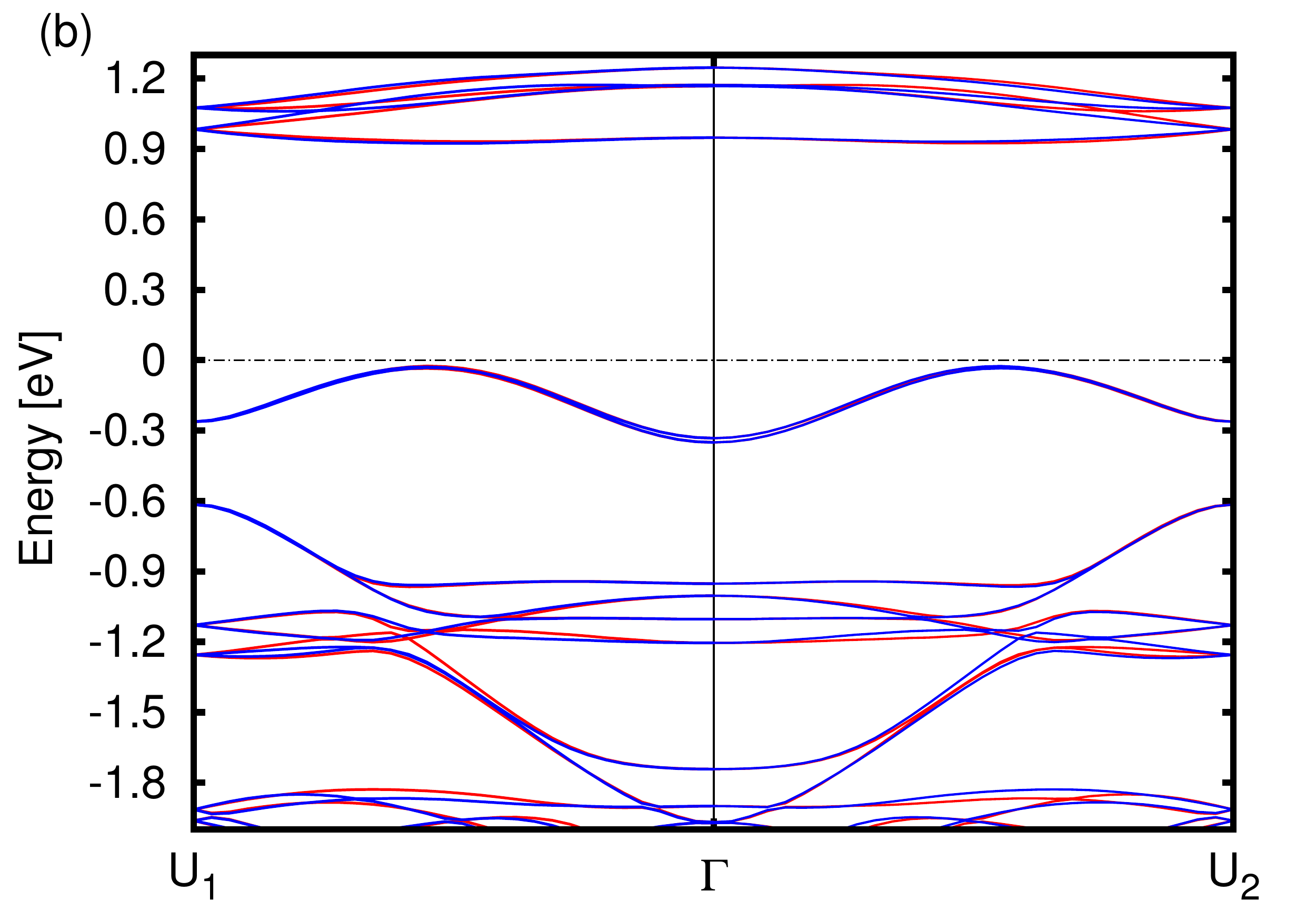}
\includegraphics[width=6.19cm,angle=270]{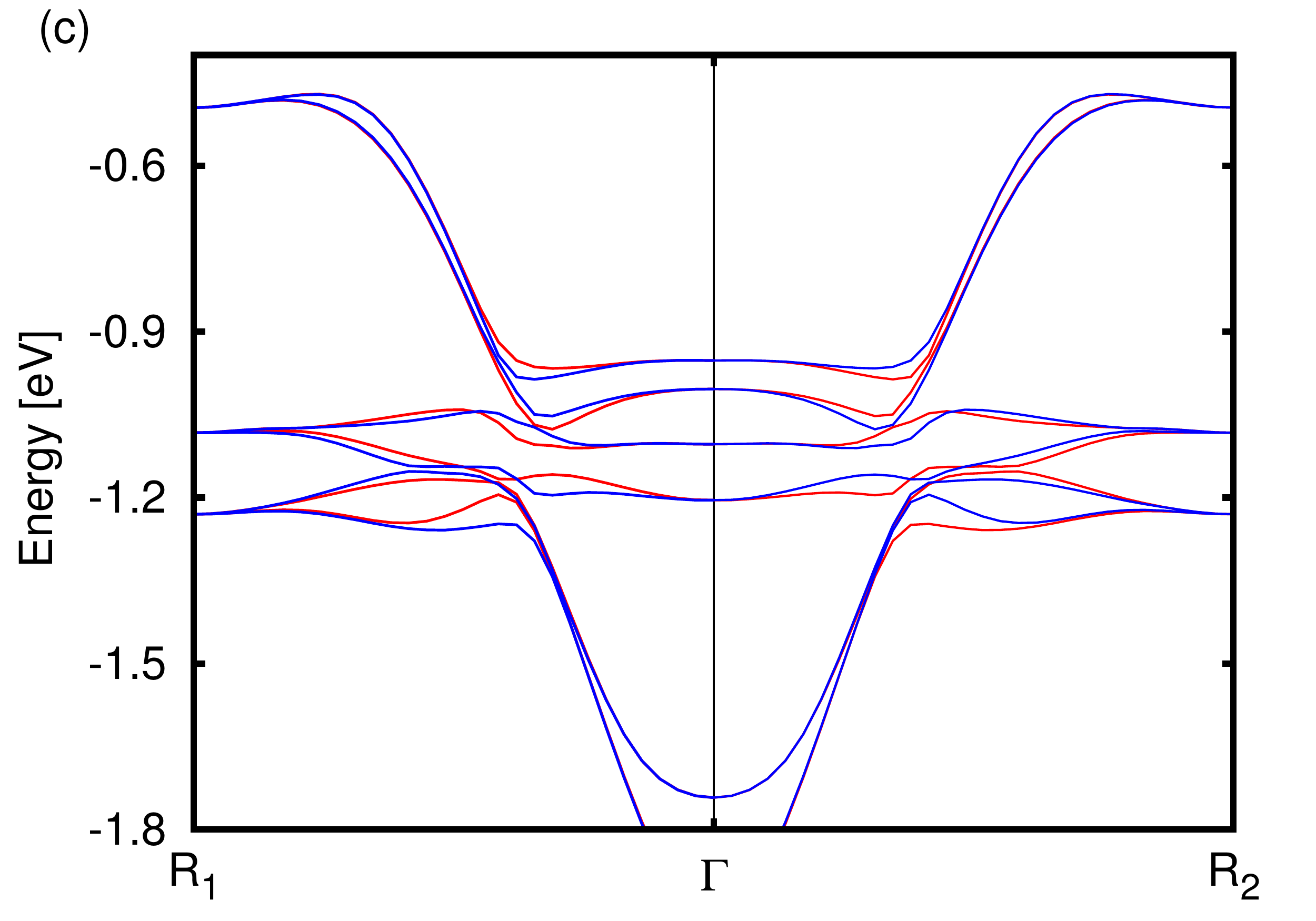}
\caption{(a) Band structure of Ca$_2$RuO$_4$ along the high-symmetry positions R$_1$-$\Gamma$-R$_2$. 
(b) Band structure of Ca$_2$RuO$_4$ along the high-symmetry positions U$_1$-$\Gamma$-U$_2$.
(c) Magnification of the band structure along the R$_1$-$\Gamma$-R$_2$ k-path with energy between -1.8 and -0.4 eV.
Blue and red lines represent the spin-up and spin-down channels, respectively.}\label{BS_CAO}
\end{figure}

In the strongly-correlated transition metal oxides, the interplay between magnetism and orbital physics plays an important role\cite{Brzezicki_2020}.
Looking for altermagnetism with large non-relativistic spin-splitting, we need to search for systems with strong orbital order. The orbital-order could be in the e$_g$ subsector\cite{Autieri_2014,AUTIERI2023414407,PhysRevB.89.155109} or in the t$_{2g}$ subsector. In this paper, we focus on compounds with orbital order in the latter. 
To investigate the altermagnetism in transition metal oxides, we choose two systems belonging to the two major classes of oxide perovskites:
the quasi two-dimensional A$_2$BO$_4$ and three-dimensional ABO$_3$. 
Regarding the quasi two-dimensional case, we  
investigate Ca$_2$RuO$_4$ while for the three-dimensional case, we study YVO$_3$.
In both cases, the interplay between the electronic correlation and the orbital physics is extremely relevant in the t$_{2g}$ subsector, which dominates the low energy physics. 
Ca$_2$RuO$_4$ is an antiferromagnetic Mott insulator with orbital-selective Mott physics \cite{PhysRevLett.104.226401,doi:10.1021/acs.nanolett.3c00574,curcio2023current} and spin-orbital correlation driven negative thermal expansion\cite{PhysRevB.107.104403}. Recently, the metal-insulator transition induced by electric current or electric field was deeply investigated in this compound\cite{PhysRevB.100.235142}.
The large family of vanadate oxides AVO$_3$ (A = La,Y,...) was intensively studied for the interplay between orbital and magnetic orders and the metal-insulator transition present in this material class\cite{PhysRevLett.99.126402,PhysRevB.103.035129,BRZEZICKI2022168616,PhysRevB.87.045132,PhysRevLett.122.127206}. Recently, a member of the vanadates family has been predicted to be a rare Kugel-Khomskii system where a purely electronic mechanism can drive the orbital-order via the superexchange\cite{PhysRevB.106.115110}. We decide to investigate YVO$_3$ which presents the largest distortions and therefore the most robust orbital-order among the members of this family.


Using first principle calculations, we demonstrate the presence of altermagnetism in both Ca$_2$RuO$_4$ and YVO$_3$. We discover an orbital-selective altermagnetism in Ca$_2$RuO$_4$ and we report the evolution of the altermagnetism as a function of the magnetic order and Coulomb repulsion in the vanadates family.


The paper is organized as follows: in the second Section, we illustrate the computational details and symmetries of the systems. In the third Section, we study the orbital-selective altermagnetism in quasi two-dimensional Ca$_2$RuO$_4$ while in the fourth Section, we show our results for the correlation-enhanced non-relativistic spin-splitting in YVO$_3$. Finally, we draw our conclusions.

\section{Computational details and symmetries}

We performed density functional theory calculations by using the VASP package\cite{Kresse93,Kresse96,Kresse96b}. All the calculations have been done without considering relativistic effects. 
For the Ca$_2$RuO$_4$, we have investigated the so-called short phase S-Pbca with space group no. 61. We have used a Coulomb repulsion of U=3 eV on the d-orbitals of the Ru-4d\cite{Autieri_2016}, the remaining computational details for the Ca$_2$RuO$_4$ were reported in a previous paper\cite{ma15196657}.
For the YVO$_3$, a plane-wave energy cut-off of 400~eV has been used. As an exchange-correlation functional, the generalised gradient approximation (GGA) of Perdrew, Burke, and Ernzerhof has been adopted\cite{Perdew96}. We mapped the momentum space with a 8$\times$8$\times$6 $k$-mesh centered at $\Gamma$ with 100 independent $k$-points in the Brillouin zone (BZ). We have used the Coulomb repulsion U on the 3d-V orbitals with the Hund coupling J$_H$=0.15U.
We have scanned U between 0 and 5 eV with the best value being around U=3 eV\cite{KUMARI201720}.
The crystal structure is the Pbnm with space group no. 62 observed upon cooling below 77 K. We have used the experimental positions and the lattice constants reported at 5 K\cite{Reehuis06}.

\section{Orbital-selective altermagnetism in C\lowercase{a}$_2$R\lowercase{u}O$_4$}

\begin{figure}[t!]
\centering
\includegraphics[width=3.99cm,angle=0]{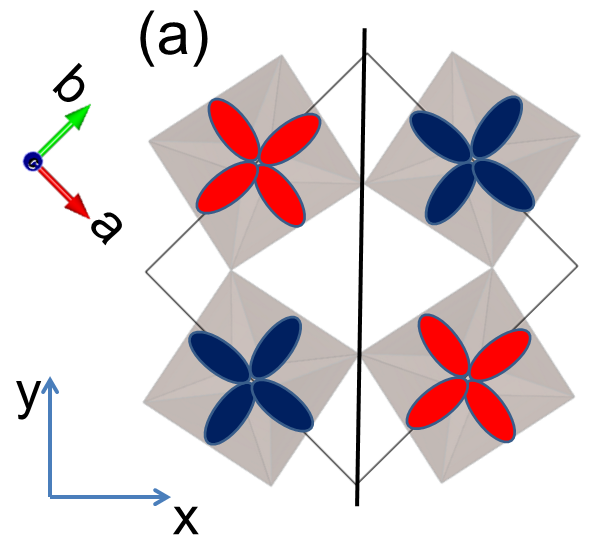}
\includegraphics[width=3.99cm,angle=0]{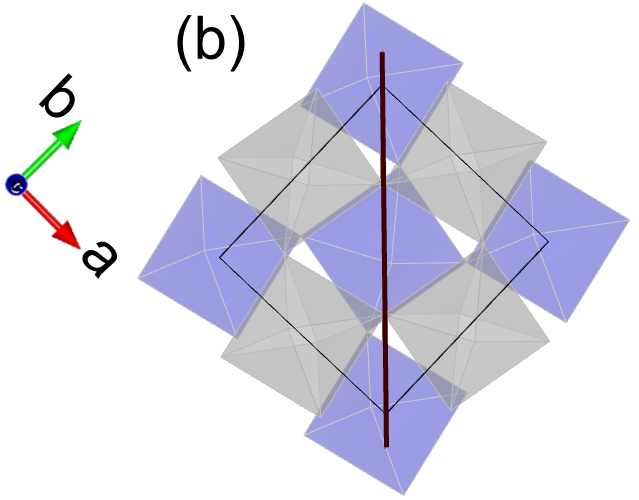}
\includegraphics[width=3.99cm,angle=0]{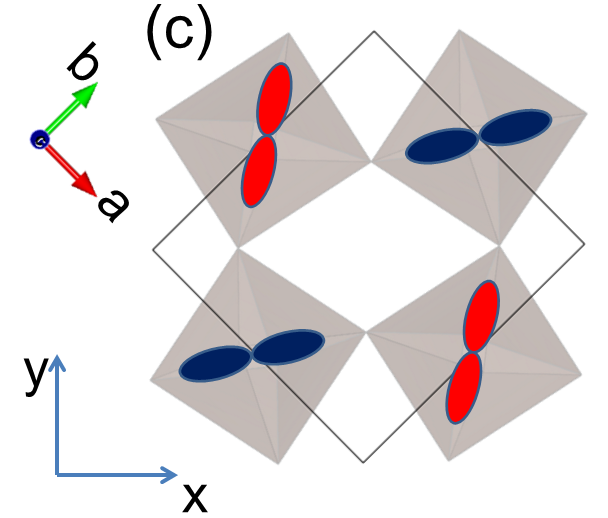}
\includegraphics[width=3.99cm,angle=0]{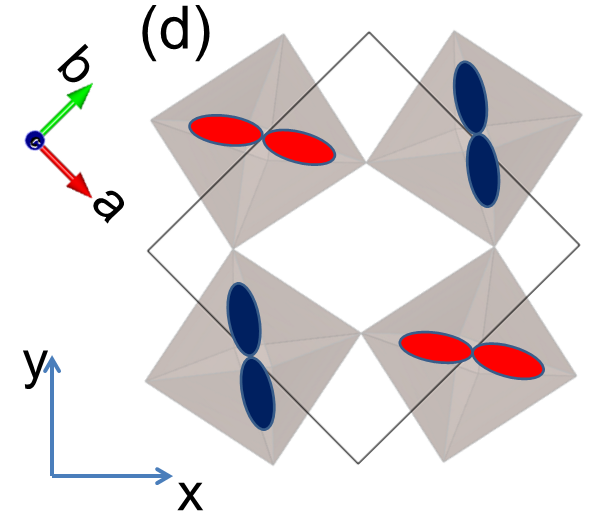}
\caption{(a) Schematic top view of the d$_{xy}$ majority electrons in a 2D plane cut by a perpendicular mirror plane in black. (b) The same plane used as a mirror in the 2D case is not a mirror plane anymore in the 3D crystal structure of Ca$_2$RuO$_4$. The grey and light blue colors indicate the octahedra in different planes. (c) Schematic top view of d$_{xz}$ spin-up and d$_{yz}$ spin-down majority electrons. (d) Schematic top view of d$_{yz}$ spin-up and d$_{xz}$ spin-down majority electrons. We use the color blue and red for the spin sublattices up and down, respectively.}
\label{Schematic_CAO}
\end{figure}

The symmetries of the BZ for the antiferromagnetic phase of Ca$_2$RuO$_4$ are reported in Fig. \ref{Brillouin_CAO}. 
With the subscripts 1 and 2, we indicate the two points in the k-space that have opposite non-relativistic spin-splitting. 
We plot the non-relativistic band structure of Ca$_2$RuO$_4$ in Fig. \ref{BS_CAO} along two different $k$-paths R$_1$-$\Gamma$-R$_2$ (top panel) and U$_1$-$\Gamma$-U$_2$ (middle panel). 
In Fig. \ref{BS_CAO}(c) we show a magnification of the band structure along R$_1$-$\Gamma$-R$_2$.
These k-paths represent the regions of the $k$-space where the altermagnetic spin-splitting is maximum, but it survives in all the BZ except for the high-symmetry directions where one of the \textit{k}-component is zero or on the zone boundaries of the BZ.
The band structures are composed of two sets of bands. The first set is composed of 4 weekly dispersive bands at around -1 eV and 4 weekly dispersive bands at around +1 eV with respect to the Fermi level. The other set is more dispersive and it is composed of 2 bands in the top of the valence and the other 2 bands that have a bandwidth from -0.5 to -2.0 eV. The less dispersive bands at around -1 eV and +1 eV have mainly d$_{xz}$/d$_{yz}$ character, while the others are mainly d$_{xy}$. 
The lowest d$_{xy}$ band has a bandwidth between -0.5 and -2.0 eV while it hybridizes with the majority d$_{xz}$/d$_{yz}$ bands around -1 eV. 
In both k-paths, we observe the presence of non-relativistic spin-splitting in the d$_{xz}$/d$_{yz}$ bands while the splitting is suppressed in the d$_{xy}$ bands. This orbital-selective altermagnetism cannot be explained only with orbital-dependent spin-splitting. The altermagnetic spin-splitting does depend on the spin-splitting between the majority (maj) and minority (min) spin-channels E$^{min}$-E$^{maj}$.
This quantity shows strong orbital dependence in Ca$_2$RuO$_4$. At the $\Gamma$ point, E$^{min}_{d_{xy}}$-E$^{maj}_{d_{xy}}$= 1.6 eV while  E$^{min}_{d_{xz}/d_{yz}}$-E$^{maj}_{d_{xz}/d_{yz}}$= 2.2 eV. Nonetheless, this difference is not enough to explain this large orbital-selective effect. We show that this scenario is made possible by the quasi-two-dimensionality of a system with t$_{2g}$ electrons.
The d$_{xy}$ orbitals have negligible hopping parameters between the layers, therefore we can assume that the d$_{xy}$ bands are two-dimensional \cite{PhysRevB.74.035115,PhysRevB.85.075126}. Therefore, we have a mirror plane that produces an approximated mirror symmetry and suppresses the altermagnetism as we can see in the schematic picture Fig. \ref{Schematic_CAO}(a). However, the d$_{xz}$/d$_{yz}$ orbitals are 3D and this plane is not a mirror anymore when we go in 3D as we can see in Fig. \ref{Schematic_CAO}(b). 
In Fig. \ref{Schematic_CAO}(c,d), we report the top view of the d$_{xz}$/d$_{yz}$ orbitals. We observe that there is no mirror for the 3D orbitals but there is a rotation of 90 degrees that map the d$_{xz}$/d$_{yz}$ spin-up in the d$_{yz}$/d$_{xz}$ spin-down and vice-versa realising the altermagnetism.

\begin{figure*}[t!]
\centering
\includegraphics[width=0.32\linewidth]{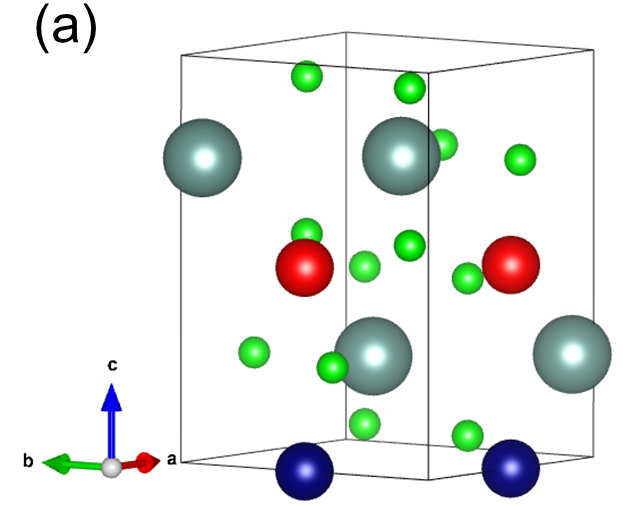}
\includegraphics[width=0.32\linewidth]{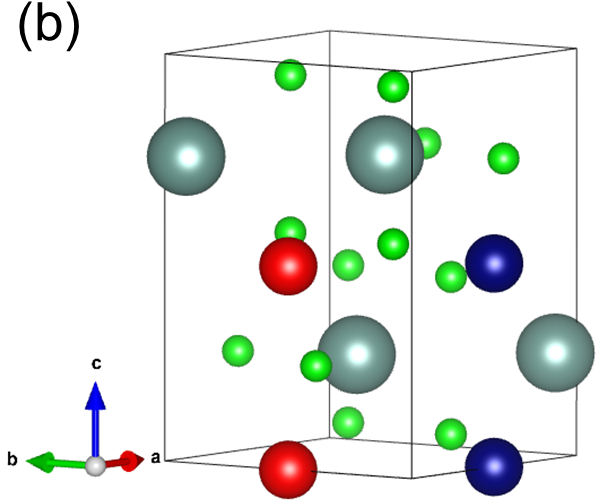}
\includegraphics[width=0.32\linewidth]{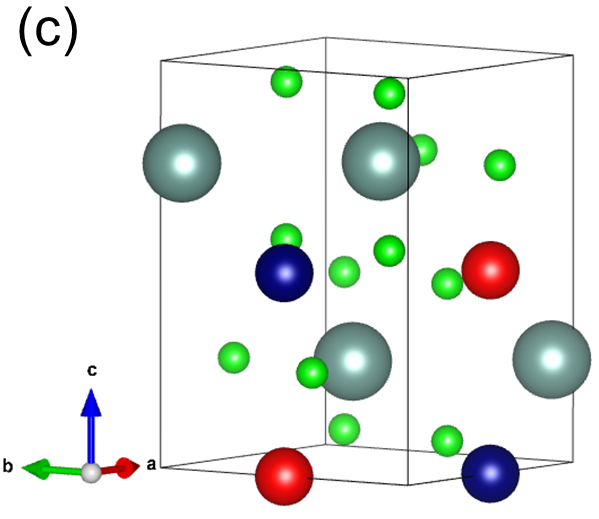}
\caption{Crystal structure of YVO$_3$ for the (a) A-type, (b) C-type and (c) G-type magnetic order in the Pbnm setting. The spheres with colors blue and red represent the V atoms with the opposite spin moments. The grey spheres are the Y atoms, while the green spheres are the O atoms. The magnetic space groups for A-type, C-type and G-type configurations are described in the text.}\label{Crystal_structure}
\end{figure*}

\begin{figure*}[t!]
\centering
\includegraphics[width=1\linewidth]{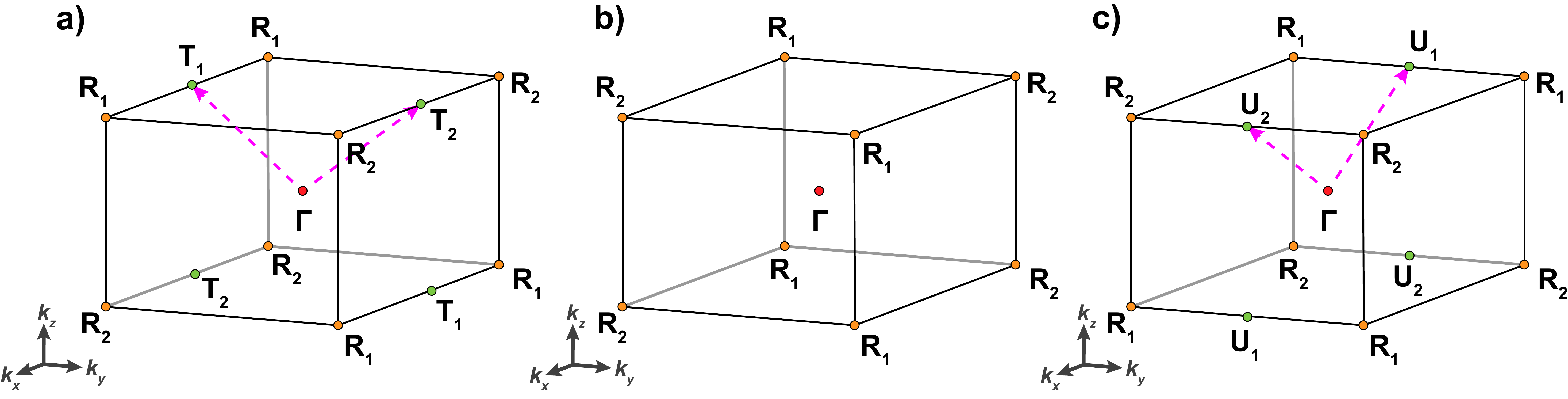}
\caption{Symmetries of the Brillouin zone for the (a) A-type, (b) C-type and (c) G-type magnetic order of YVO$_3$ (space group no. 62) in the Pbnm setting. In our notation, the high-symmetry points with the subscript 1 and 2 show altermagnetism along the path towards the $\Gamma$ point. The positions of the points R$_1$ and R$_2$ evolve as a function of the magnetic order. In A-type and G-type magnetic orders, the k-paths $\Gamma$-T and $\Gamma$-U show altermagnetic spin-splitting, respectively.}\label{Brillouin_YVO}
\end{figure*}

The AHE was deeply investigated in ferromagnetic ruthenates oxides and its interfaces for the tunable properties deriving from the Berry phase and for the presence of humps\cite{Groenendijk20,Vanthiel21,Yang21,Malsch20,Wysocki20}. Due to the orbital selective altermagnetism, we obtain that the AHE can be measured in Ca$_2$RuO$_4$ just in n-doped samples while the highest valence bands show no sizeable altermagnetism.

\section{Correlation effects on the altermagnetic properties of YVO$_3$}

\begin{figure*}[t!]
\centering
\includegraphics[width=4.7cm,height=5.9cm,angle=270]{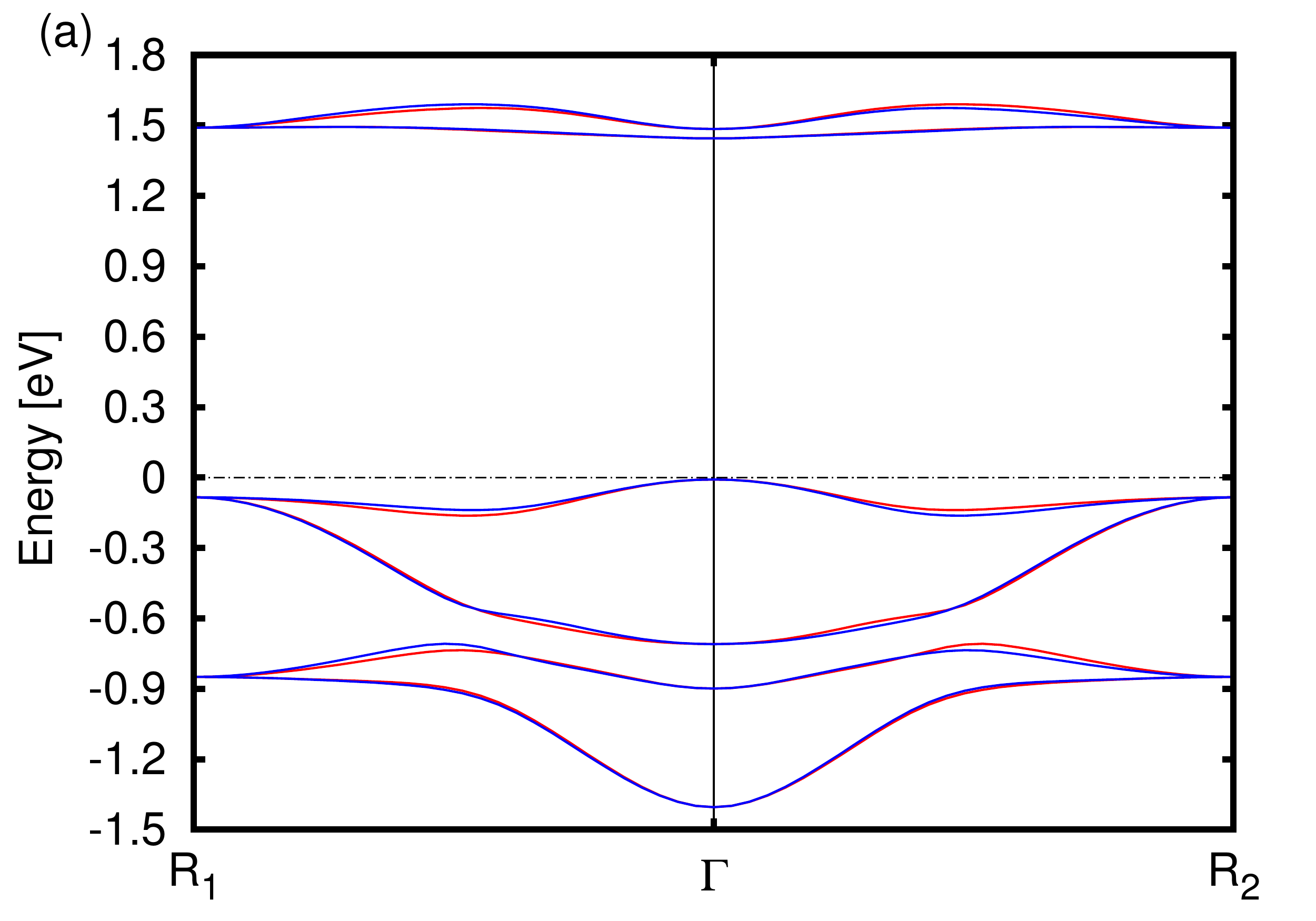}
\includegraphics[width=4.7cm,height=5.9cm,angle=270]{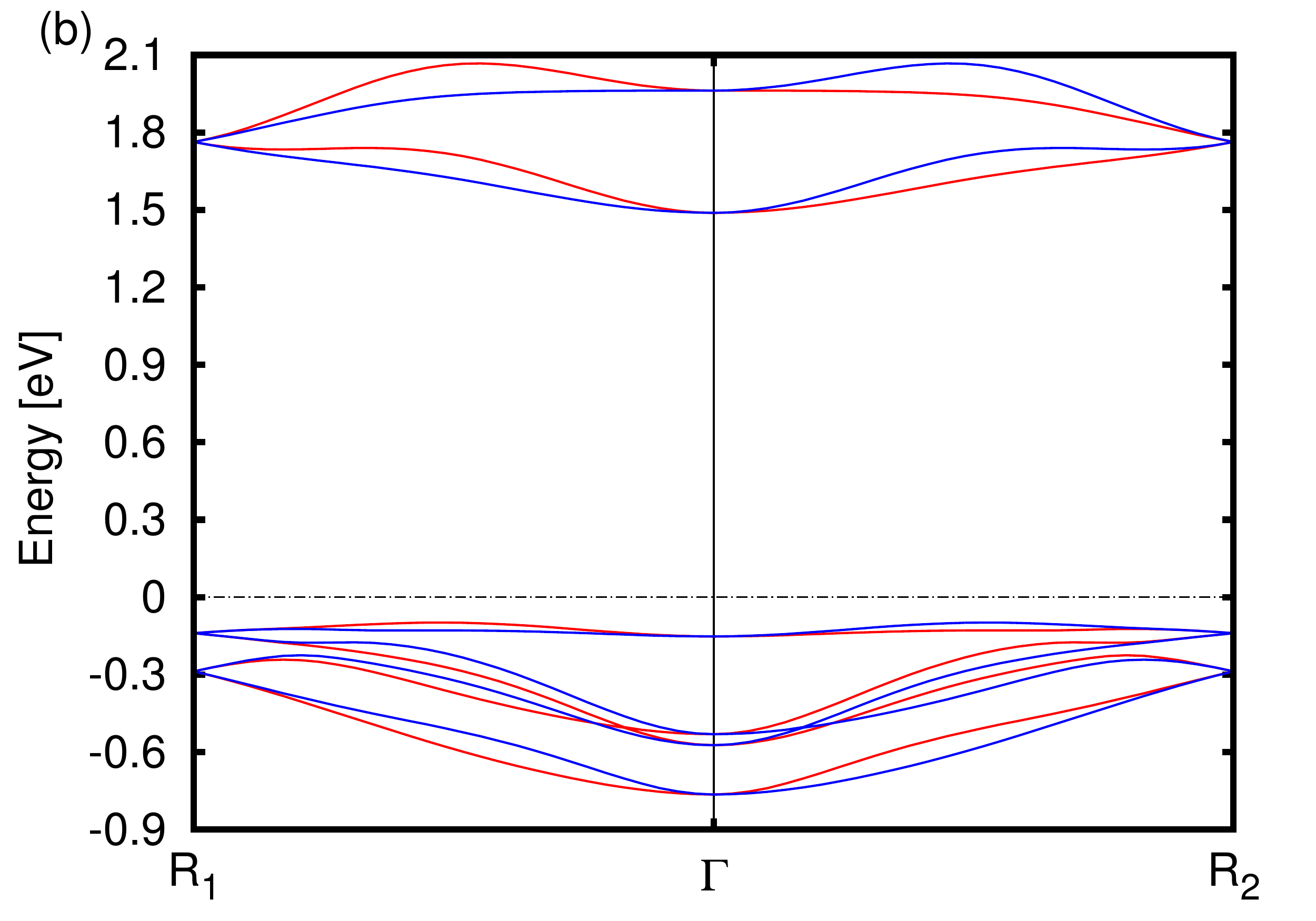}
\includegraphics[width=4.7cm,height=5.9cm,angle=270]{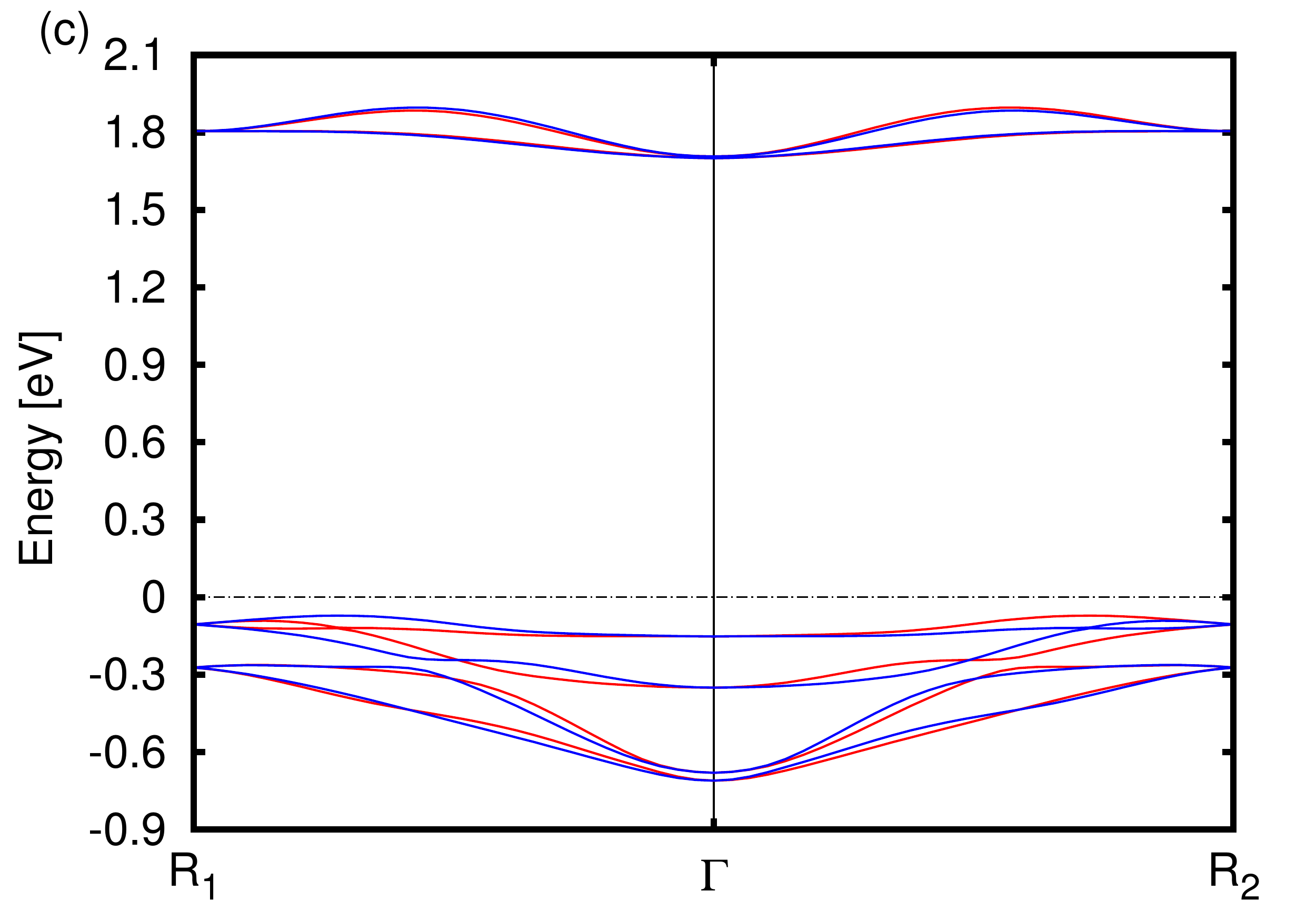}
\caption{Spin-splitting along the  $R_1$-$\Gamma$-$R_2$ for the majority t$_{2g}$ V-bands of YVO$_3$ for the (a) A-type, (b) C-type and (c) G-type magnetic order. In these cases, we have used the value of U= 3 eV.  The spin-up channel is shown in blue, while the spin-down channel is shown in red.}\label{BS_YV}
\end{figure*}

Now we shift our attention to the second class of oxide perovskites that we want to investigate, namely the three-dimensional ABO$_3$.
We analyze the altermagnetism in YVO$_3$ as a function of the magnetic order. The magnetic orders usually investigated in transition-metal perovskite ABO$_3$ with four formula units in the unit cell are A-type, C-type and G-type, shown in Fig. \ref{Crystal_structure} together with the magnetic space group numbers and types\cite{Gallego:db5106}. The band gaps of the three different magnetic orders at U=3 eV are 1.39 eV, 1.48 eV  and 1.69 eV, respectively. The altermagnetism is present in all magnetic orders and this outcome is not granted by the space group No. 62.
Indeed, by changing the magnetic configuration, the system changes the magnetic space group. Since YVO$_3$ stays altermagnetic in all magnetic configurations, it means that all the magnetic space groups of the different configurations belong to types I and III.
Paying attention to consider the correct setting for the crystal structure, the magnetic space groups for A-type are 62.448, 62.447 and 62.441 for the N\'eel vector along x, y and z, respectively. The C-type configurations belong to the space groups 62.446, 62.441 and 62.447 for the N\'eel vector along x, y and z, respectively, while for the G-type configuration, the magnetic space groups are 62.441, 62.446 and 62.448 for the N\'eel vector along x, y and z, respectively. The space group 62.441 belongs to Type-I while all  others belong to the type-III magnetic space groups.
In other compounds with space group No. 62 but different crystal structures and magnetic space group, the altermagnetism was observed just in one of the magnetic orders \cite{Fakhredine23}.
In Fig. \ref{Brillouin_YVO}, we show the way how symmetries of the BZ evolve which is not straightforward as a function of the magnetic order. For the C-type magnetic order, we have a checkerboard pattern of R$_1$ and R$_2$ while no altermagnetism is present neither along $\Gamma$-U nor $\Gamma$-T as shown in Fig. \ref{Brillouin_YVO}(b). For the A-type and G-type, we have consecutive R$_1$ and R$_2$ along the k$_x$ and k$_y$ axis, respectively. The altermagnetism is present also along the $\Gamma$-T and $\Gamma$-U paths for the A-type and G-type magnetic orders, respectively. As a consequence, the C-type magnetic order has fewer regions of the k-space where altermagnetism is present. 
We report in Fig. \ref{BS_YV} the band structure along the R$_1$-$\Gamma$-R$_2$ path for the three magnetic orders. Despite the minor region of the k-space with altermagnetism, the C-type is the magnetic order that shows larger non-relativistic spin-splittings. The A-type shows very small non-relativistic spin-splittings, while in the G-type magnetic order we obtain intermediate values for the spin-splittings.

\begin{figure}[t!]
\centering
\includegraphics[width=6.8cm,height=8.8cm,angle=270]{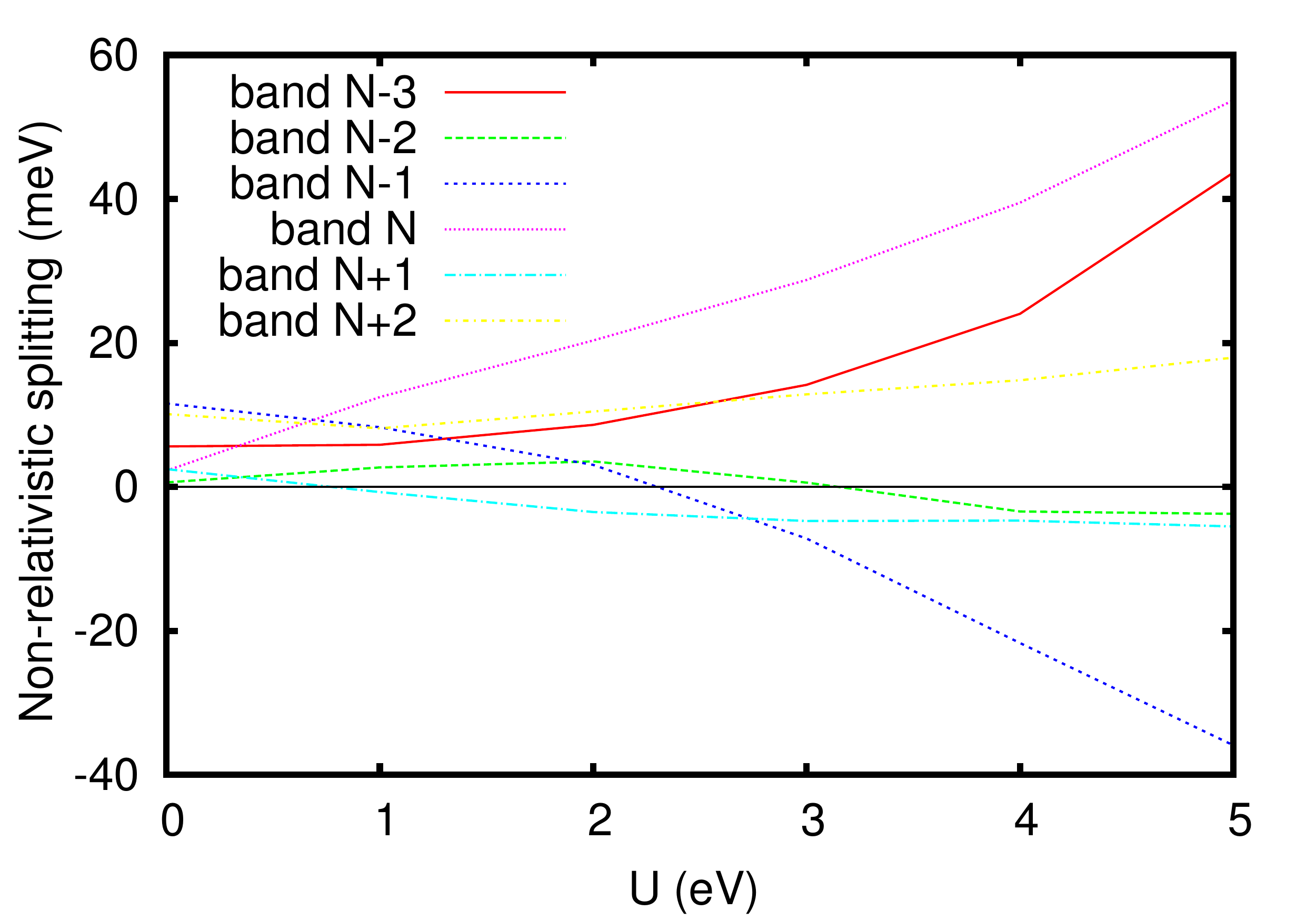}
\caption{Evolution of the spin-splitting at the vector k=(0.25,0.25,0.25) as a function of the Coulomb repulsion for all 6 bands of the majority t$_{2g}$ V-orbitals. The band number N is the highest valence band while the band number N+1 is the lowest conduction band. We considered the absolute value of the splitting at U=0 and we assume it as negative when the spin-splitting changes sign.}\label{Splitting_YV}
\end{figure}

Finally, we report the evolution of the non-relativistic spin-splitting as a function of the Coulomb repulsion for the majority t$_{2g}$ V-orbitals. 
We consider the midpoint between $\Gamma$ and R in the momentum space, since the large spin-splitting is zero at $\Gamma$ and R and it is usually maximum at halfway\cite{Smejkal22,Smejkal22beyond}. The Coulomb repulsion is uniform over the Brillouin zone, therefore it will affect the band structure at all k-points in the same way. The absolute values of the spin-splittings for all six bands are considered at U=0, if the spin-splitting change sign we use the negative value. The results are reported in Fig. \ref{Splitting_YV} for the G-type magnetic order. 
From the plot, we conclude that the spin-splitting at U=0 is relatively small due to the weak electronic dispersion of the strongly-correlated bands. However, the Coulomb repulsion enhances on average the non-relativistic spin-splitting. Indeed, we have a correlation-enhanced non-relativistic spin-splitting up to a factor 4-5 with respect to the initial splitting at U=0. The motivation for this large increase is the following. The relevant quantity for the altermagnetic spin-splitting is the spin-splitting between minority and majority t$_{2g}$ electrons E$^{min}_{t_{2g}}$-E$^{maj}_{t_{2g}}$, this quantity increases with U. As the splitting between minority and majority increases with U, the altermagnetic spin-splitting on average follows a similar trend. 

As a weak point of this analysis, in both Ca$_2$RuO$_4$ and YVO$_3$ the non-relativistic spin-splitting is relatively weak despite the strong orbital-order in YVO$_3$. Therefore, to obtain very large non-relativistic spin-splitting we need both strong orbital-order and large hopping parameters. The latter are usually not present in strongly-correlated systems.

\section{Conclusions}

Using \textit{first-principles} calculations, we have investigated the altermagnetism in strongly correlated systems with an orbital order in the t$_{2g}$ subsector, choosing  two systems belonging to the two major classes of oxide perovskites:
the quasi two-dimensional A$_2$BO$_4$ and three-dimensional ABO$_3$. We have found an orbital-selective altermagnetism in quasi 2D Ca$_2$RuO$_4$. While the d$_{xz}$/d$_{yz}$ orbitals are connected only by a roto-translation as requested condition to observe the altermagnetism, the quasi-twodimensionality of the d$_{xy}$ bands allows introducing a mirror plane that strongly suppresses the altermagnetism in the d$_{xy}$ bands. 
In the YVO$_3$, the altermagnetism is present in all magnetic configurations. We have shown how the symmetries of Brillouin zone evolve in the three-dimensional YVO$_3$ as a function of the magnetic order. The non-relativistic spin-splitting is larger for the C-type magnetic order. We report that the Coulomb repulsion enhances the non-relativistic spin-splitting making the strongly-correlated system a platform for the research of altermagnetic properties.\\

\begin{acknowledgments}
We thank Tomasz Dietl, Mario Cuoco, Adolfo Avella and Amar Fakhredine for useful discussions.
The work is supported by the Foundation for Polish Science through the International Research Agendas program co-financed by the European Union within the Smart Growth Operational Programme (Grant No. MAB/2017/1).
J. S. is supported by the National Science Centre (Poland) OPUS
2021/41/B/ST3/04475. 
We acknowledge the access to the computing facilities of the Interdisciplinary Center of Modeling at the University of Warsaw, Grant G84-0, GB84-1 and GB84-7. We acknowledge the CINECA award under the ISCRA initiative  IsC85 "TOPMOST" and IsC93 "RATIO" grant, for the availability of high-performance computing resources and support. We acknowledge the access to the computing facilities of the Poznan Supercomputing and Networking Center Grant No. 609.
\end{acknowledgments}

\bibliography{altermagnetism}
\end{document}